# Seafloor crustal deformation data along the subduction zones around Japan obtained by GNSS-A observations

Yusuke Yokota[1], Tadashi Ishikawa[1] and Shun-ichi Watanabe[1]

**Affiliations**
1. Hydrographic and Oceanographic Department, Japan Coast Guard, 3-1-1 Kasumigaseki, Chiyoda-ku, Tokyo 100-8932, Japan
corresponding author(s): Yusuke Yokota (eisei@jodc.go.jp)

## Abstract
**Crustal deformation data obtained by geodetic observation networks are foundations in the fields of geodesy and seismology. These data are essential for understanding plate motion and earthquake sources and for simulating earthquake and tsunami disasters. Although relatively scarce, seafloor geodetic data are particularly important for monitoring the behaviour of undersea interplate boundary regions. Since the mid-1990s, we have been developing the combined Global Navigation Satellite System-Acoustic ranging (GNSS-A) technique for realizing seafloor geodesy. This technique allows us to collect time series of seafloor crustal deformation. Our published data can be used to investigate several seismological phenomena along the subduction zones around Japan, namely the Nankai Trough, Sagami Trough and Japan Trench. These regions are globally important places in geodesy and seismology and are also suitable for comparison with other geophysical datasets. Our intention is for these data to promote further understanding of megathrust zones.**

## Background & Summary

Japanese Islands are located in a tectonically active region where multiple tectonic plates interact with each other. Interplate megathrust earthquakes have occurred many times along the undersea plate subduction zone, causing serious damage to human society. To prevent further disasters, it is important to elucidate the physical mechanisms of such earthquakes. Crustal deformation data derived from geodetic observation networks, e.g., Global Navigation Satellite System (GNSS) and INterferometric Synthetic-Aperture Radar (InSAR), are extremely useful for revealing the inter-seismic slip-deficit distributions and the pre-, co- and post-seismic events. In Japan, the Geospatial Information Authority of Japan (GSI) has established a dense GNSS observation network called GEONET (GNSS Earth Observation Network System)[1]. GEONET has detected interplate earthquake cycles and various slow slip events occurring on interplate boundaries[2-3].

However, despite the fact that interplate megathrust earthquakes occur in undersea subduction zones, most crustal deformation data to date have been collected by onshore networks only. This lack of data from marine regions limits the estimation resolution and reliability for undersea interplate slip and slip-deficit. Therefore, there is increasing demand for seafloor crustal deformation data with which to elucidate interplate megathrust earthquakes.

Satellite geodesy using electromagnetic waves (e.g. GNSS, Very-Long-Baseline Interferometry (VLBI) and Satellite Laser Ranging (SLR)) enables absolute onshore position to be determined precisely. However, electromagnetic waves cannot be used to determine absolute seafloor position (SP) because of scattering and absorption by seawater. As proposed originally by ref. 4, the combined GNSS-Acoustic ranging (GNSS-A) technique determines SP through a combination of radio-wave GNSS data from above the sea and acoustic-wave ranging data from under the sea. Since the mid-1990s, the Hydrographic and Oceanographic Department of the Japan Coast Guard (JHOD) has been developing a seafloor geodetic observation system based on this technique[5] and has been constructing a seafloor geodetic observation network along the Nankai Trough, Sagami Trough and Japan Trench. The location of our seafloor observation site is shown in Fig. 1. The onsite measurements with this system are made aboard a survey vessel, on a campaign basis. Table 1 shows the list of campaign epoch used in this work. The data described here are coordinates of SP for each campaign epoch. The time series of SP so obtained represent seafloor crustal deformation. These are mainly due to phenomena around the plate boundary. For example, steady trend of the time series mainly reflect deformation due to subduction of the oceanic plate. A step-like change is due to a co-seismic deformation. These data are strongly meaningful for understanding subduction zones and the processes of megathrust earthquakes.

Co-seismic deformations due to some megathrust earthquakes have been recorded in these data. The 2005 Off-Miyagi Prefecture Earthquake (Mw 7.2) and the 2011 Tohoku-oki Earthquake (Mw 9.0) were clearly recorded by our observation network[6-7]. Their post-seismic deformations were also recorded[8-9]. The data following the 2005 Off-Miyagi Prefecture Earthquake also include a process preceding the Tohoku-oki Earthquake[8,10]. In this way, the behaviours of earthquake cycles along the Japan Trench were recorded collectively just above the source regions. These data have been used to discuss the complicated processes involved in the earthquake cycles[10-14] and could also be used for simulations to predict future subduction processes.

Ref. 15 suggested that our network had recorded the inter-seismic coupling process along the Sagami Trough. Ref. 16 showed inter-seismic velocity fields in the Nankai Trough megathrust zone and calculated a slip-deficit rate distribution that suggested the interplate coupling condition. These data and results could also be used to discuss and simulate earthquake cycles, future tsunamis and future compound disasters.

In this paper, we present a body of time-series data accumulated since 2001 with description of our observation method and data processing. These data widely will further the understanding of not only megathrust earthquakes and earthquake disasters but also other geoscience regions.

## Methods

### Observation System

Fig. 2 shows the GNSS-A seafloor geodetic observation system. It consists of a seafloor unit with multiple acoustic mirror transponders and a sea surface unit with an acoustic transducer, a GNSS antenna-receiver and a dynamic motion sensor. The sea surface equipment is set on a surface vehicle such as a survey vessel or a buoy; our group has been using a survey vessel only.

The position of the GNSS antenna is determined continuously at a sampling rate of 2 Hz by baseline kinematic GNSS analysis using 'IT' (Interferometric Translocation) software[17-18]. The positions of the onshore GEONET reference stations used in the baseline analysis are determined by the daily coordinates of the GEONET F3 solutions[19]. In addition to the GNSS observations, the dynamic motion sensor records the attitude of the survey vessel at a sampling rate of 50 Hz. By combining the position of the GNSS antenna and the attitude of the survey vessel, the position of the underwater acoustic transducer according to the coordinate system of the GEONET F3 solutions is determined. The distance between the sea surface transducer and each seafloor transponder is determined by acoustic ranging. The transponder receives an acoustic signal from the sea surface transducer and sends it back. The round-trip travel time is determined by cross correlation. In each observation epoch, we acquire acoustic travel time data from several hundred shots for each transponder. In the data processing described in the next subsection, the absolute SP is determined by combining the data regarding absolute sea surface transducer position and acoustic travel time.

In this process, each travel time is transformed into a distance between the sea surface and seafloor units using an ocean sound speed structure (SSS) model. To construct the SSS model, sound speed profiles are acquired every few hours using temperature and salinity profilers, namely a conductivity–temperature–depth (CTD) profiler, an expendable conductivity–temperature–depth (XCTD) profiler and expendable bathythermographs (XBT).

Before July 2002, the sea surface system used a rigid 8-m-long aluminium alloy pole mounted at the stern of the vessel with a GNSS antenna at the top and an acoustic transducer at the bottom, as shown in Fig. 2a. With the pole system, the measurements were possible only when the vessel was drifting in order to avoid noise from the propellers, but they were affected by deformation of the pole by water flow. Because the vessel was drifting, the track lines were uncontrollable with this system; ideally, track lines should be spatially well balanced for improved accuracy. This situation is similar to the dilution of precision (DOP) of GNSS positioning due to bad satellite geometry. In addition, it requires time for the transfer of the vessel between track lines. Because human operations on the deck are necessary for lifting up and down the acoustic transducer for the vessel transfer, we could not conduct observations at night for safety reasons. For these reasons, it took several days to obtain the empirically required number of data. Differences in the equipment at each period were written in the data file as the equipment code. For the data from this period, the equipment code written in the data file is 'A'.

In July 2002, the sea surface system began using a more-rigid stainless steel pole to reduce the effects of pole deformation due to water flow. For the data from this period, the equipment code is 'B'.

In April 2008, the sea surface system was mounted on the vessel permanently. An acoustic transducer was mounted on the hull of the vessel together with a GNSS antenna mounted on the top of the main mast[20], as shown in Fig. 2b. With the hull-mounted system, measurements are conducted while the vessel sails along ideally well-balanced track lines at a speed of 5–8 knots. With this rigging, the measurement accuracy was improved. In addition, measurements are conducted continuously without being interrupted by the vessel transfer between track lines and human operations on the deck are no longer necessary. Consequently, the observation time was decreased from 2–4 days (when drifting) to 16–24 hours (when sailing). However, because multiple vessels were used after this period, there are differences in instrument errors between the vessels. For the data from this period, the equipment code is 'C'. The history of these changes is listed in Table 2.

**Data Processing**

The SP is determined using linearized inversion based on the least-squares formulation, combining the obtained data. The scheme is essentially that constructed by ref. 21.

The basic construction of this inversion can be represented using the general formulation of an observation equation using an *n*-dimensional observation vector $Y$ and an *m*-dimensional true model parameter vector $X$:

$$Y = f(X) + e, \tag{1}$$

where $e$ is the observation error vector that varies around zero in a Gaussian distribution characterized by the variance–covariance matrix $E$. To estimate via linearized inversion, this usually nonlinear relationship should be transformed to a linear formulation. Using small perturbations $x$ around initial value $x_0$, Equation (1) can be replaced with

$$Y = f(x_0 + x) + e. \tag{2}$$

When $x$ is sufficiently small, the right-hand side of Equation (2) can be rewritten as a Taylor expansion:

$$Y = f(x_0) + \frac{\partial f}{\partial x}x + \frac{\partial^2 f}{\partial x^2}x^2 + \cdots + e. \tag{3}$$

By ignoring terms of order $x^2$ and higher, we obtain the linear observation equation

$$y = Ax + e, \tag{4}$$

where $y \equiv Y - f(x_0)$ stands for the residuals between observation value and that calculated from the initial condition and $A$ is the $n \times m$ matrix

$$A = \begin{bmatrix} \frac{\partial f_1}{\partial x_1} & \cdots & \frac{\partial f_1}{\partial x_m} \\ \vdots & \ddots & \vdots \\ \frac{\partial f_n}{\partial x_1} & \cdots & \frac{\partial f_n}{\partial x_m} \end{bmatrix}. \tag{5}$$

The most probable $\hat{x}$ is obtained by least squares as follows:

$$\hat{x} = (A^T E^{-1} A)^{-1}(A^T E^{-1} y). \tag{6}$$

We obtain the final solution by using an iterative numerical calculation in which $x_0$ is modified slightly. The variance–covariance matrix $C$ for the resultant $\hat{x}$ is given as

$$C = (A^T E^{-1} A)^{-1}. \tag{7}$$

To estimate the SP accurately, the SSS must be sufficiently accurate. Because of the complex spatial and temporal variations of the SSS, it is impossible in practice to make enough observations to cover all these variations in detail. Thus, for centimetre-level positioning, it is not sufficient to use an observed sound speed data only. The observed travel-time data include not only information about the range but also about the SSS along the ray path. Taking advantage of this fact, we estimate an SSS correction function. This idea is similar to estimating the atmospheric delay in a precise GNSS data analysis.

The basic flow of our inversion strategy consists of two parts. First, as indicated in (I), inversion is performed to determine the SPs using a certain SSS. Then, using the resultant SPs, an SSS correction function is estimated. We iterate this process until the SPs converge (see Fig. 3).

**(I) Inversion for SP**

When we estimate the SP, the model parameter $x$ is given by three components of small corrections to the initial SP coordinates as follows:

$$x = (\Delta x \ \Delta y \ \Delta z)^T. \tag{8}$$

The observation vector $y$ is given by

$$y = (\Delta T_1 \ \Delta T_2 \ \ldots \ \Delta T_q)^T, \tag{9}$$

where $q$ is the total number of travel-time data and $\Delta T_i$ is the $j$th residual between the observed travel time and that calculated from the initial position of seafloor transponder and observed position of sea surface transducer. The travel time is calculated via ray tracing using a given SSS.

At each site, seafloor unit usually consists of four transponders array that are set approximately on a circle of radius equal to the depth. If the each transponder's positions are determined independently, the array geometry at each epoch changes due to the various error sources at each epoch. However, the array geometry of the seafloor transponders (in a 1–3-km square) is not distorted in about one or two decades generally.

For highly accurate estimation, each epoch's displacements are determined under the assumption that the array geometry is not distorted and array geometry is estimated using all the datasets. This method was developed by ref. 22. To determine the array geometry using all the datasets, the observation equation is represented as follows:

$$\begin{pmatrix} y_1 \\ y_2 \\ y_3 \\ \vdots \end{pmatrix} = \begin{pmatrix} A_1 & 0 & 0 & \cdots \\ A_2 & A_2 & 0 & \cdots \\ A_3 & 0 & A_3 & \cdots \\ \vdots & \vdots & \vdots & \ddots \end{pmatrix} \begin{pmatrix} X_1 \\ D_2 \\ D_3 \\ \vdots \end{pmatrix} + e, \tag{13}$$

where $X_1$, the epoch-1 position of each transponder, is a model parameter for the array geometry, 3-dimensional vector $D_j$ is a model parameter for the epoch-$j$ displacement of the array from epoch-1 position, $y_i$ is an observation vector of the epoch-$i$ and $A_i$ is a matrix of partial derivatives of the epoch-$i$. The SP of the epoch-$i$ is represented as $X_i = X_1 + D_i$.

Array is replaced at periodic intervals due to battery restrictions. Since simultaneous parallel observation among old and new transponders is performed after replacing, the same reference point can be decided from either old or new transponder group as shown in Fig. 4. The provided position is given as the time-series of the same reference point which is a centre of all the transponders installed in the past.

**(II) Inversion to correct sound-speed structure**

The SSS correction is represented by a certain function of time. The correction function $\Delta V(t)$ can be represented by a linear combination of basis functions $\Phi_i(t)$:

$$\Delta V(t) = \sum_k a_k \Phi_k(t), \tag{10}$$

where $a_k$ is the coefficient of the $k$th basis function. The model parameter vector $x$ in Equation (1) is given by the coefficients of the basis functions as follows:

$$x = (a_1 \; a_2 \; a_3 \; a_4 \ldots)^T. \tag{11}$$

To obtain the geometric path of an acoustic wave, we used ray tracing under the horizontally layered SSS. In general, the sound speed varies with each layer: the near-surface layer is more variable than the abyssal one, for example, so the correction function must be estimated for each layer. However, the geometric path of an acoustic wave is almost a straight line because of the smallness of the refractive index. For this reason, the positioning result depends mainly on the average sound speed of all layers. In this analysis, $\Delta V$ is not set for each layer but is set uniformly for all the layers.

This process removes the effects of those SSS disturbances that contain long-period components (e.g. daily variation) and short-period components. In many cases, the short-period components include spatial variations due to the vessel's movements.

## Data Records

Our published data are in the form of a time series of the centre position of the seafloor transponder array derived from the above processing. Because we use the GEONET F3 solutions as reference positions in our GNSS analyses, the coordinates of our data are consistent with the International Terrestrial Reference Frame 2005 (ITRF2005)[23]. The time series shows the absolute positions as the Earth-Centred Earth-Fixed (ECEF) coordinates.

The data file also includes metadata, a list of which is given in Table 3. Date of updating, site name, reference coordinate system and each observation record are written in the file. Each observation record consists of observation start date, absolute positions [XYZ coordinates], equipment and vessel codes and observation end date. Since one observation may span multiple days, observation start and end days are written. The name of the survey vessel used in each epoch is also written in the data file to distinguish instrument errors, as described in the next section. Four vessels have been used. Codes T, S, M and K are allocated to survey vessels of Takuyo, Shoyo, Meiyo and Kaiyo, respectively. The data files are distributed in text format via the PANGAEA (Data Citation 1). End of the file name is '_xyz.txt'. To see crustal deformations, it is useful to transform the ECEF coordinates into local East, North and Up (ENU) coordinates. We also provide the ENU data. The time series shows the relative movements from the first epoch in ENU direction. End of the file name is '_enu.txt'. Metadata in this file is written in Table 4.

The site along the Japan Trench had greatly moved before and after the 2011 Tohoku-oki earthquake[7]. The array shape had changed and at the same time some transponders became unusable. Therefore, the data files are separated at the period of March 2011. Provided Tohoku-oki earthquake co-seismic movement data of the sites along the Japan Trench were calculated using only transponders that can be compared before and after the earthquake in the same way as ref. 7 ('Tohoku_EQ_Movements_enu/xyz.txt', Data Citation 1). Site names, periods, vessel codes and movements consistent with ITRF2005 are written in these data. As above, since the time series before and after the Tohoku-oki earthquake were not observed using the same array, the difference between these time series cannot be calculated using the provided time series data.

Because arrays at the sites along the Nankai and Sagami Troughs were not distorted or broken, the 2011 Tohoku-oki earthquake co-seismic movement data is not provided for these sites. The co-seismic effects due to the earthquake are included in the time series data of these sites.

## Technical Validation

The provided data are affected by three main sources of error (Fig. 2c)[25-26]. The first is the influence of the atmosphere, ionosphere and others on the GNSS positioning (GNSS error). The second is due to the vessel equipment, such as the relative position error between the GNSS antenna and the on-board acoustic transducer (G-T error). There are also instrumental errors in a motion sensor and a transducer. In order to distinguish these instrumental errors which should be different for each vessel, identifiers of vessels are allocated for each epoch. The third is an error from the SSS that cannot be removed sufficiently by the data processing (SSS error). These errors have the potential to act not only as random errors (which would be reduced by increasing the amount of acoustic-wave ranging data) but also as systematic errors (which would not be reduced by increasing the amount of acoustic-wave ranging data). The variances of the model parameters as indicated by the diagonal components of $C$ in Equation (7) are considered to reflect the random errors. However, in many cases the calculated variances are much smaller than the scatter of the final SP time series. This indicates that the influence of the systematic errors cannot be ignored but also cannot be evaluated statistically (e.g. random-walk noise in the high-rate GNSS analysis and spatial heterogeneity of the SSS). It is difficult to evaluate individual systematic errors quantitatively. For example, because it is also impossible to know the true position of the vessel on the sea at every moment, we cannot evaluate the deviation of the estimated position from the true one. Therefore, our scheme cannot provide substantive information about the positioning error at each epoch.

Instead, we recommend evaluating the entire data set statistically, for example deriving the average crustal deformation speed by regression analysis. While the uncertainty of each epoch cannot be calculated, as described above, the SP repeatability can be evaluated by deviation from the regression line because a surface of the ground generally deforms linearly. Since the crustal deformation is believed to have been stable for a long time unless large earthquakes occur, we can consider the time-series regression line to be the maximum likelihood estimate. Therefore, any dispersion of the data from the line is deemed an indicator of positioning uncertainty. The uncertainty of the velocity vector derived from the trend of the regression line can be calculated by the ordinary mathematical method of regression analysis.

As a concrete example of time series, eastward movement of site MYGI are shown in Fig. 5. Two large earthquakes occurred near site MYGI during observation period. To avoid co-seismic steps and nonlinear post-seismic movements, linear regression was applied in three periods. Root mean square (RMS) of the residuals from regression line which is an indicator of positioning uncertainty was 2.7 cm, 3.0 cm and 1.5 cm, respectively.

The published data should not be evaluated using only two epochs before and after the event because the dispersion in the data is larger than the assumed crustal deformation in many cases. When considering a co-seismic step, we recommend evaluating the data of several epochs before and after an event statistically. For such evaluation, it should be noted that steady crustal movements cannot be ignored because of the sparse observation frequency. For example, co-seismic movement associated with the 2005 event is obtained from a difference between two regression lines. However the obtained value (3.1 cm) is almost the same level as RMS. Therefore according to statistical testing, there were not enough significant differences.

For estimate the positioning uncertainty of our system, we investigated the residuals from the linear regression lines of all the data after 2013. Fig. 6 shows a horizontal scatterplot and histograms for three components that indicate the residuals for all sites. Horizontal scatterplot shows that almost data are not separated by 5 cm from the regression lines. Comparing the histograms, the states of residuals of eastward and northward components are similar in the value of variances and the tail of distribution. On the other hands, the histogram for vertical components has larger variances value and shorter tail. According to the *F*-test when the significant level is set to 0.05, the hypothesis that the variances of vertical component have equal to ones of horizontal components is rejected, while the hypothesis that the variances of horizontal components are equal is not rejected.

The total observation uncertainty of this technique is estimated empirically as approximately 2 cm (RMS) in the horizontal component for each epoch. The uncertainty in the vertical SP component is greater than that in the horizontal one because we observe only the upper region of the seafloor. This situation is similar to that for GNSS positioning. Additionally, correlation between the vertical SP component and the SSS reduces the accuracy. There is a trade-off between the vertical SP component and the SSS. For example, if the given sound speed is higher than the true value, a deeper SP is estimated without an appreciable increase in the residuals.

## Acknowledgements
We thank O. L. Colombo of the NASA Goddard Space Flight Center for providing us with the kinematic GNSS software 'IT' (Interferometric Translocation), and we thank the Geospatial Information Authority of Japan (GSI) for providing us with the high-rate GNSS data for the kinematic GNSS analysis and the daily coordinates of the GEONET sites. Many JHOD staff, including the crews of the S/Vs *Takuyo*, *Shoyo*, *Meiyo* and *Kaiyo*, supported our observations and data processing.

## Author contributions
TI and YY conceived and designed the study. TI and YY wrote the manuscript. TI, SW and YY discussed the contents and approved the final manuscript.

# Tables

Table 1 | List of seafloor sites and observation epochs. Upper and bottom tables indicate the periods before and after the 2011 Tohoku-oki earthquake, respectively.

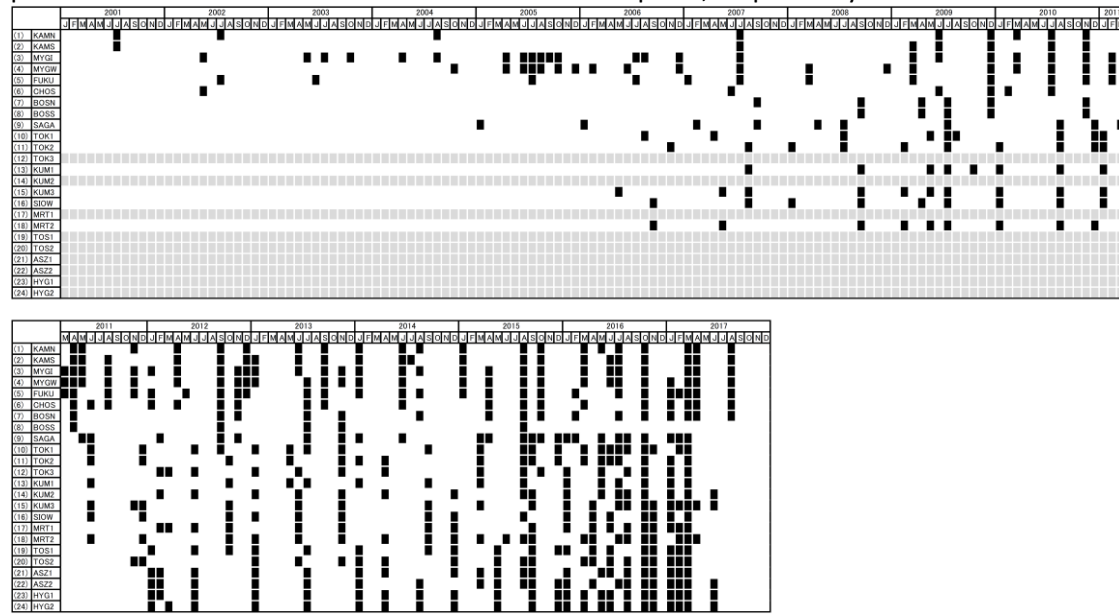

Table 2 | History of changes in equipment and methodology for each survey vessel.

|   | Date | Transducer system | Material of pole |   |
|---|---|---|---|---|
| A | ~ Jul. 2002 | Pole system | Aluminium alloy | One epoch was calculated using several days' observation. |
| B | ~ Apr. 2008 | Pole system | Stainless |  |
| C | -- | Hull-mounted | -- | There is an instrument error for each vessel. |

Table 3 | Published XYZ data format including metadata.

| Line number | Contents |
|---|---|
| 1st line | Date of updating |
| 2nd line | Site name; reference coordinate system |
| 3rd line | Description of each column |
| after 4th line | Observation start date; absolute positions [XYZ coordinates]; equipment code; vessel code; Observation end date |

Table 4 | Published ENU data format including metadata.

| Line number | Contents |
|---|---|
| 1st line | Date of updating |
| 2nd line | Site name; reference coordinate system |
| 3rd line | Origin of the plane E-N-U coordinate system (latitude, longitude, ellipsoidal height) |
| 4th line | Position of reference epoch with (E, N, U = 0.0, 0.0, 0.0) from origin point described in 3rd line |
| after 5th line | Movements from the reference epoch |

## Data Citation

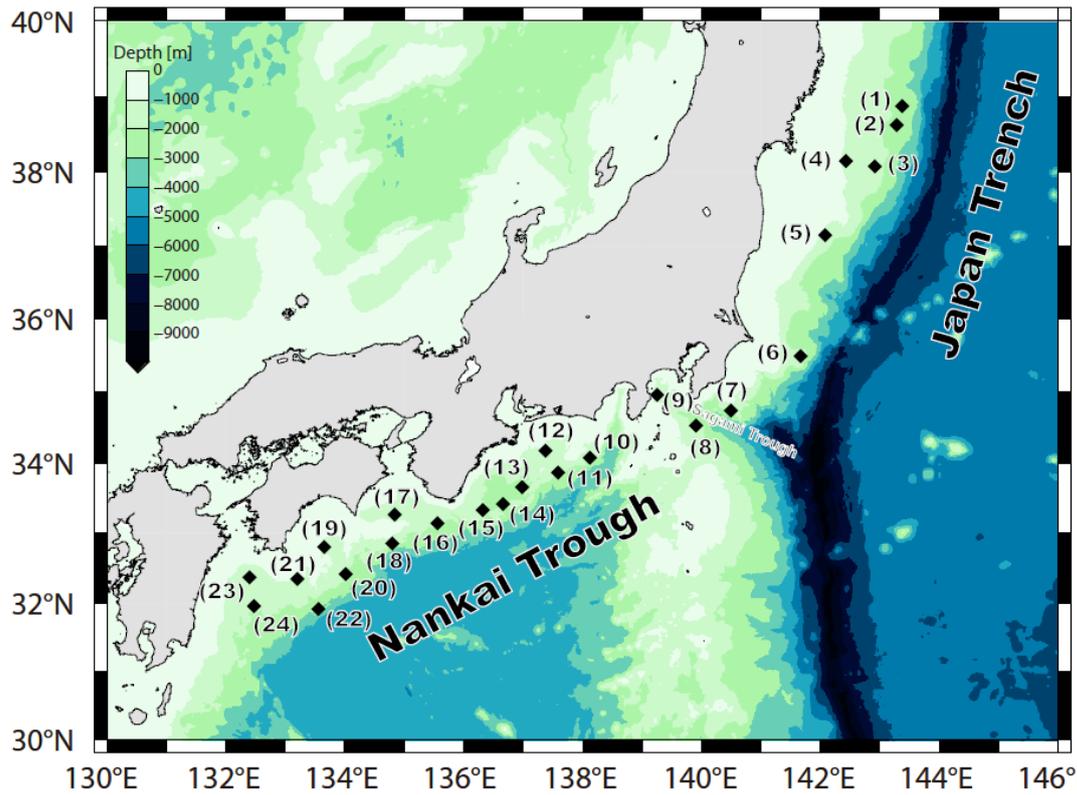

Figure 1 | Installed positions of seafloor geodetic observation sites. Site names are described in Table 1.

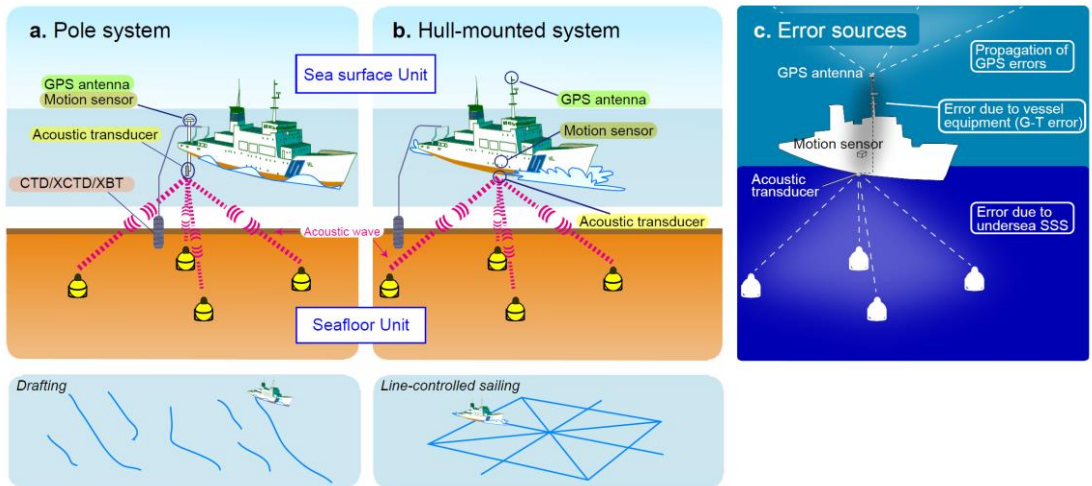

Figure 2 | Schematics of our seafloor geodetic observation using **a.** pole system and **b.** hull-mounted system. **c.** Schematic of error sources. This figure was modified from refs. 5, 7, 15, 16, 21 and 26.

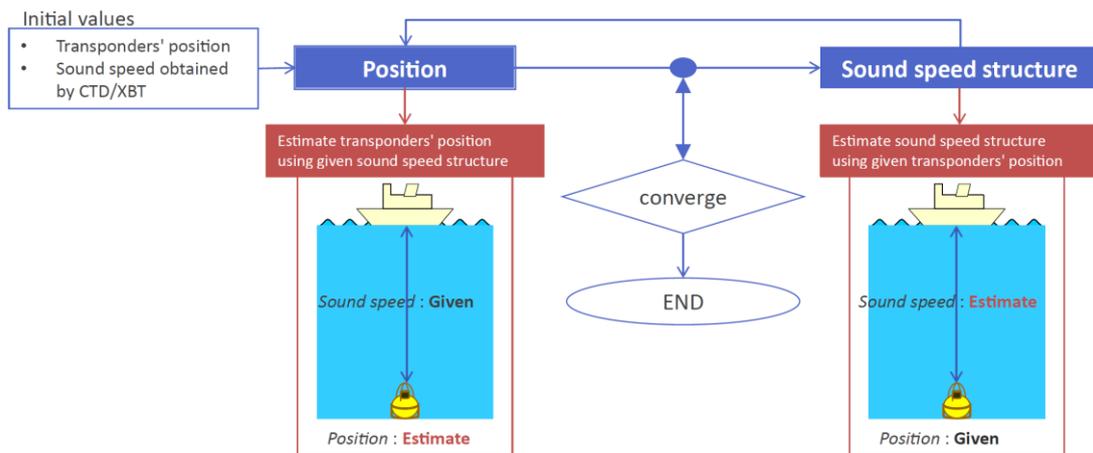

Figure 3 | Algorithm flow of data processing applied to obtain the absolute seafloor position using transducer position data and acoustic data. This figure was modified from ref. 21.

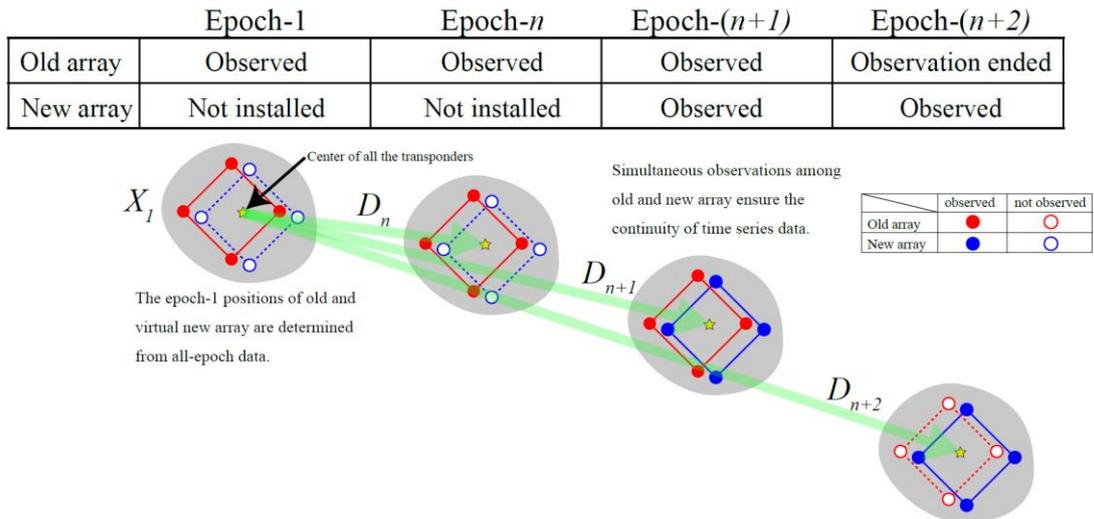

Figure 4 | Schematics about replacing a transponder. Yellow stars indicate positions of center of all the transponders. The time series of yellow stars are provided in our dataset. Red and blue circles indicate the installed and replaced transponders, respectively. Outlined circles are transponders that are not observed in the epochs.

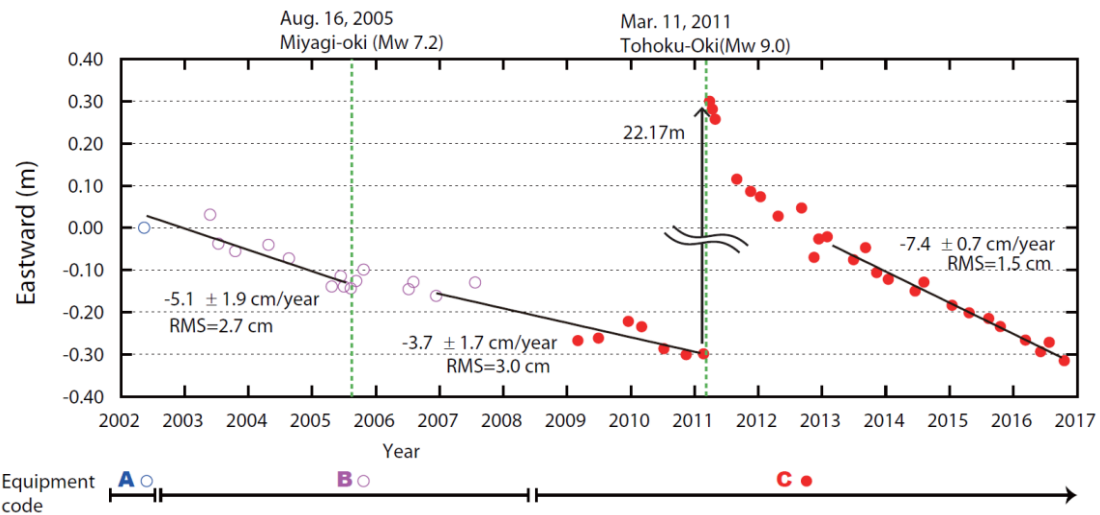

Figure 5 | An example of time series of site MYGI. The reference frame is ITRF2005[23]. Blue, purple and red circles indicate data in the periods of equipment codes A, B and C, respectively. Black lines are the linear regression lines for the periods before the 2005 Off-Miyagi Prefecture earthquake, before the 2011 Tohoku-oki earthquake and after 2013, respectively. Green dashed line show the 2005 and the 2011 earthquakes.

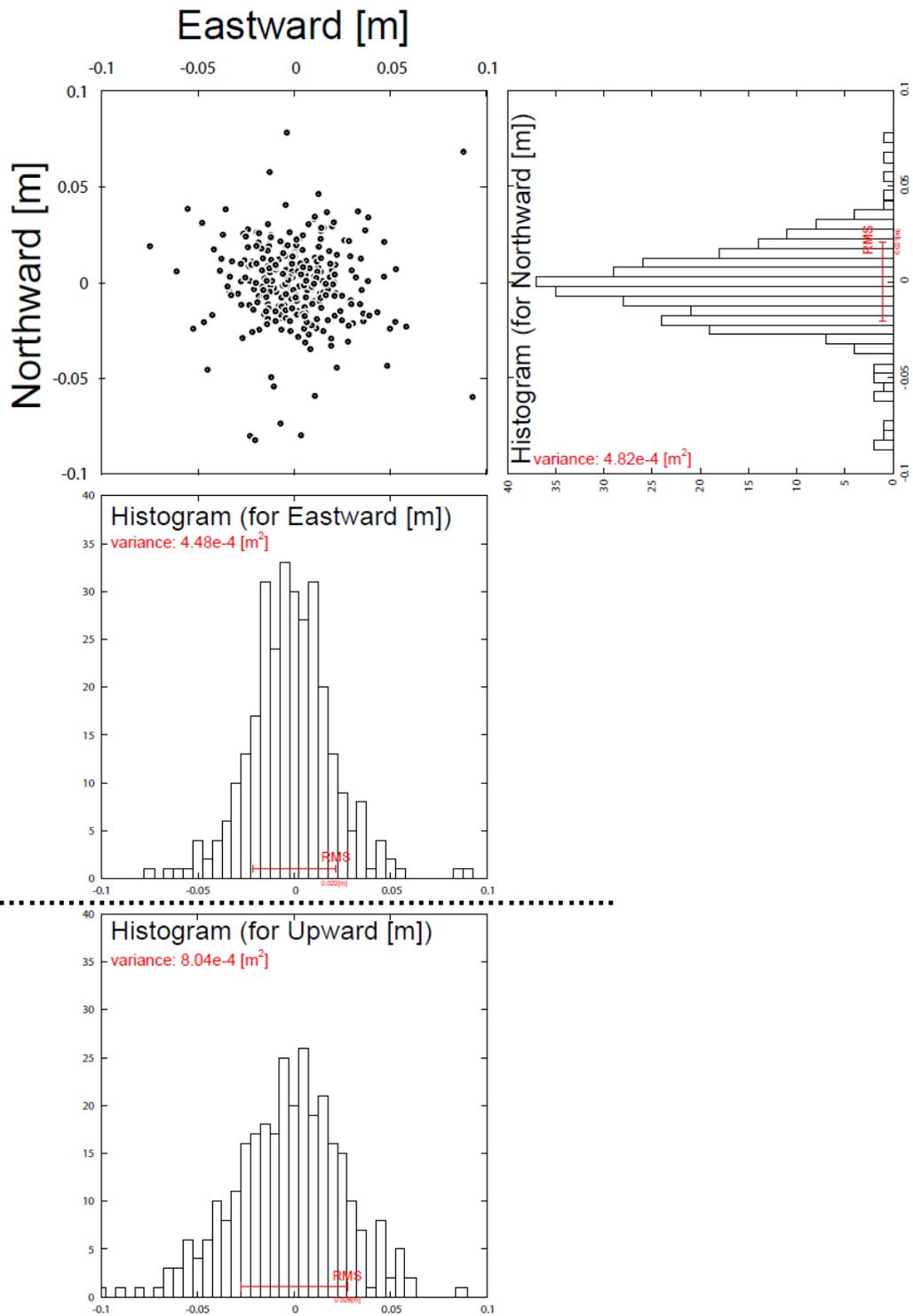

Figure 6 | Scatterplot of residuals from the linear regression lines of all the data after 2013. Histograms on the right and below show dispersions in the northward, eastward and upward components, respectively.